\documentclass[amsmath,twocolumn,superscriptaddress,10pt,aps,pra]{revtex4-1}%
\usepackage{amsmath}%
\usepackage{amsfonts}%
\usepackage{amssymb}%
\usepackage{graphicx}
\usepackage{fullpage}
\usepackage{dsfont}
\usepackage{silence}
\usepackage{bm}
\usepackage{multirow}
\usepackage{color}
\usepackage{float}
\usepackage[T1]{fontenc}

\usepackage[utf8]{inputenc}

\begin{document}

\title{Microscopic modelling of exciton-polariton diffusion coefficients in atomically thin semiconductors}%in MoSe$_2$ monolayers}

\author{Beatriz Ferreira}
\email{beatriz.ferreira@chalmers.se}
\affiliation{Chalmers University of Technology, Department of Physics, 412 96 Gothenburg, Sweden}

\author{Roberto Rosati}
\affiliation{Department of Physics, Philipps-Universit\"at Marburg, Renthof 7, D-35032 Marburg, Germany}

%\author{...}
%\affiliation{...}

\author{Ermin Malic}
\affiliation{Department of Physics, Philipps-Universit\"at Marburg, Renthof 7, D-35032 Marburg, Germany}
\affiliation{Chalmers University of Technology, Department of Physics, 412 96 Gothenburg, Sweden}

\begin{abstract}
In the strong light-matter coupling regime realized e.g. by integrating semiconductors into optical microcavities,
polaritons as new hybrid light-matter quasi-particles are formed. The corresponding change in the dispersion relation has a large impact on optics, dynamics and transport behaviour of semiconductors.  
In this work, we investigate the strong-coupling regime in hBN-encapsulated MoSe$_2$ monolayers focusing on  exciton-polariton diffusion. Applying a microscopic approach based on the exciton density matrix formalism combined with the Hopfield approach, we predict a drastic increase of the diffusion coefficients by two to three orders of magnitude in the strong coupling regime. We explain this behaviour by the much larger polariton group velocity and suppressed polariton-phonon scattering channels with respect to the case of bare excitons. Our study contributes to a better microscopic understanding of polariton diffusion in atomically thin semiconductors.

\end{abstract}

\maketitle

\section{Introduction}\label{sec1}
Semiconductors integrated into microcavities can be subject to a strong light-matter coupling \cite{deng2010exciton,sanvitto2016}. In this regime, the coupling to light is stronger than the (difference of) lifetime of the involved particles, e.g. cavity modes and excitons \cite{savona1999,deng2010exciton,khitrova2006vacuum,baranov2018novel}. As a direct consequence, polaritons as dual light-matter quasi-particles are formed and considerably change the bandstructure, optics, dynamics and transport properties of semiconductors. 
A promising class of materials for the strong-coupling regime are monolayers of transition metal dichalcogenides (TMDs) \cite{liu2015,lundt2016,low2017,schneider2018,anton2021bosonic}. They exhibit a large oscillator strength and exciton binding energies in the range of a few hundreds of meV governing the optoelectronic properties of these materials \cite{he2014,ugeda2014,Wang18,Mueller18,brunetti2018} and have been hence widely investigated in planar microcavities \cite{liu2015,dufferwiel2015,schneider2018}. In the strong-coupling regime, the coupling of photons with confined excitons give rise to the formation of exciton-polaritons. The interaction leads to an avoided crossing at momenta where the cavity mode crosses the exciton dispersion resulting in a Rabi splitting, cf. Fig. \ref{fig_1}(a) \cite{dufferwiel2015,epstein2020}. Hence, the polariton dispersion is characterized by two distinct energy branches denoted as the upper and the lower polariton (UP, LP). The weight of the photonic or excitonic character of polariton branches is expressed by the Hopfield coefficients $C_\pm$ \cite{hopfield1958theory}, cf. Fig. \ref{fig_1}(b).

The quasi-bosonic polaritons show intriguing effects, such as Bose-Einstein condensation \cite{kasprzak2006bose,ma2020, anton2021bosonic}, (super)fluidity \cite{ballarini2013all, kolmakov2016toward}, topological effects \cite{klembt2018topo}, and promising applications from lasing \cite{st2017lasing}, to integrated circuits \cite{amo2010switches} and quantum computing \cite{ghosh2020quantum}. 
Exciton-polaritons maintain characteristics of both photons and excitons, such as a small effective mass, which makes them attractive also for transport purposes  \cite{steger2013,vondran2019,Wurdack21}. 
Previous studies have shown that TMD monolayers with their rich exciton landscape, including dark and bright exciton states \cite{Berghauser18,Malic18,Deilmann19}, exhibit an interesting spatio-temporal exciton dynamics resulting in an intriguing exciton diffusion behaviour. This includes non-classical diffusion \cite{Wagner21}, transient negative diffusion  \cite{Rosati20}, accelerated hot-exciton diffusion \cite{Rosati21c} or formation of spatial rings (halos) \cite{Kulig18,Perea19,Glazov19} and unconventional exciton funneling effects \cite{Rosati21d}. In view of their light component, polaritons show an interesting transport behaviour resulting in a fast propagation in the ballistic regime \cite{steger2013,Wurdack21}. 
The polariton diffusion has already been observed e.g. in ZnSe and GaAs films \cite{bley1998exciton,zaitsev2015}. 
To best of our knowledge, there have been no microscopic studies on exciton-polariton diffusion in atomically thin TMDs yet. 
In our work, we close this gap and 
present a microscopic study of polariton diffusion in an exemplary TMD monolayer integrated into an optical cavity. We 
investigate the change in the group velocity and polariton-phonon scattering rates as crucial ingredients determining the diffusion coefficient. Based on our microscopic approach, we predict a polariton diffusion coefficient up to three orders of magnitude higher compared to bare exciton diffusion. 

%%%%%%%%%%%%%%%%%%%%%%%%%%%%%%%%%%%%%%%%%%%%%%%%%%%%%%%%%%%%%%%%%%%%%%%%
%%%%%%%%%%%%%%%%%%%%%%%%%%%%%%%%%%%%%%%%%%%%%%%%%%%%%%%%%%%%%%%%%%%%%%%%
\section{Theory}\label{sec2}
First, we describe the theoretical approach allowing us to microscopically investigate  polariton diffusion coefficients in TMD monolayers. Exciton energies and wavefunctions in  TMD monolayers are obtained by solving  the Wannier equation \cite{Haug09test, berghauser2014analytical, Selig16} including DFT input on the characteristics of the electronic bandstructure \cite{Kormanyos15}. In this work, we focus on an hBN-encapsulated MoSe$_2$ monolayer, which we find to be a direct semiconductor with the bright KK excitons (electron and hole located at the K point) as energetically lowest states \cite{Malic18}.
In contrast, tungsten-based TMDs are known to be indirect semiconductors with momentum-dark excitons as  energetically lowest states \cite{Brem20,Malic18,Deilmann2019}. Since the latter are not directly affected by light-matter coupling, we expect smaller polariton-induced changes in the diffusion coefficient for tungsten-based materials.
The starting point is the many-particle Hamilton operator in excitonic picture, which reads in second quantization \cite{Katsch18test,Brem20}
\begin{equation}\label{hamilton1}
\begin{split}
      \hat{H}=&\sum_{\mathbf{Q}}E^\text{X}_{\mathbf{Q}}\hat{X}^\dagger_{\mathbf{Q}}\hat{X}_{\mathbf{Q}}+\sum_{\mathbf{Q}}E^\text{C}_{ \mathbf{Q}}\hat{c}^\dagger_{\mathbf{Q}}\hat{c}_{\mathbf{Q}}+\\
      &\sum_{\mathbf{Q}} g_{\mathbf{Q}}\left(\hat{c}^\dagger_{\mathbf{Q}}\hat{X}_{\mathbf{Q}}+\hat{c}_{\mathbf{Q}}\hat{X}^\dagger_{\mathbf{Q}}\right)+\\
  &\sum_{\mathbf{Q},\mathbf{q}}\mathcal{D}_{\mathbf{q}}\hat{X}_{\mathbf{
Q}+\mathbf{q}}\hat{X}_{\mathbf{Q}}\left(\hat{b}^\dagger_{-\mathbf{q}}+
\hat{b}_{\mathbf{q}}\right).
\end{split} 
\end{equation}
Here, $\hat{X}^\dagger_{\mathbf{Q}} (\hat{X}_{\mathbf{Q}}$),  $\hat{c}^\dagger_{\mathbf{Q}} (\hat{c}_{\mathbf{Q}}$), and $\hat{b}^\dagger_{\mathbf{q}}(
\hat{b}_{\mathbf{q}})$ are the exciton, photon, and phonon creation (annihilation) operators, respectively.  The first two terms in the Hamiltonian describe the  energy of excitons and photons with $E^\text{X}_{\mathbf{Q}}$  and $E^\text{C}_{\mathbf{Q}}$, respectively. The third term corresponds to the exciton-light interaction mediated by the exciton-photon coupling matrix element $g_{\mathbf{Q}}$, where photons need to have the same in-plane momentum $\mathbf{Q}$ as excitons to fulfill the momentum conservation. In general, the out-of plane component influences the cavity energy and  exciton-photon coupling. 
Here, we assume the existence of one resonant photon mode (i.e. $E^\text{X}_{0}=E^\text{C}_{0}$) and we study its impact on the polariton diffusion \cite{bley1998exciton}. Finally, the last term in the Hamiltonian describes the exciton-phonon interaction \cite{Selig16}, where the coupling strength is determined by the exciton-phonon matrix element $\mathcal{D}_{\mathbf{q}}$. We use DFT input parameters for phonon dispersion and the strength of the electron-phonon coupling  \cite{Jin14}. In particular, we consider three optical modes (LO, TO and A$_1$) with the energies $E_\text{TO}=36.1$ meV, $E_\text{LO}=36.6$ meV and $E_{\text{A}_1}=30.3$ meV, and for the two acoustic modes we consider a sound velocity of $v_s=4.1\times10^{-3}$ nm/fs  \cite{Jin14}. 

\begin{figure}[t!]
\centering
	\includegraphics[width=\columnwidth]{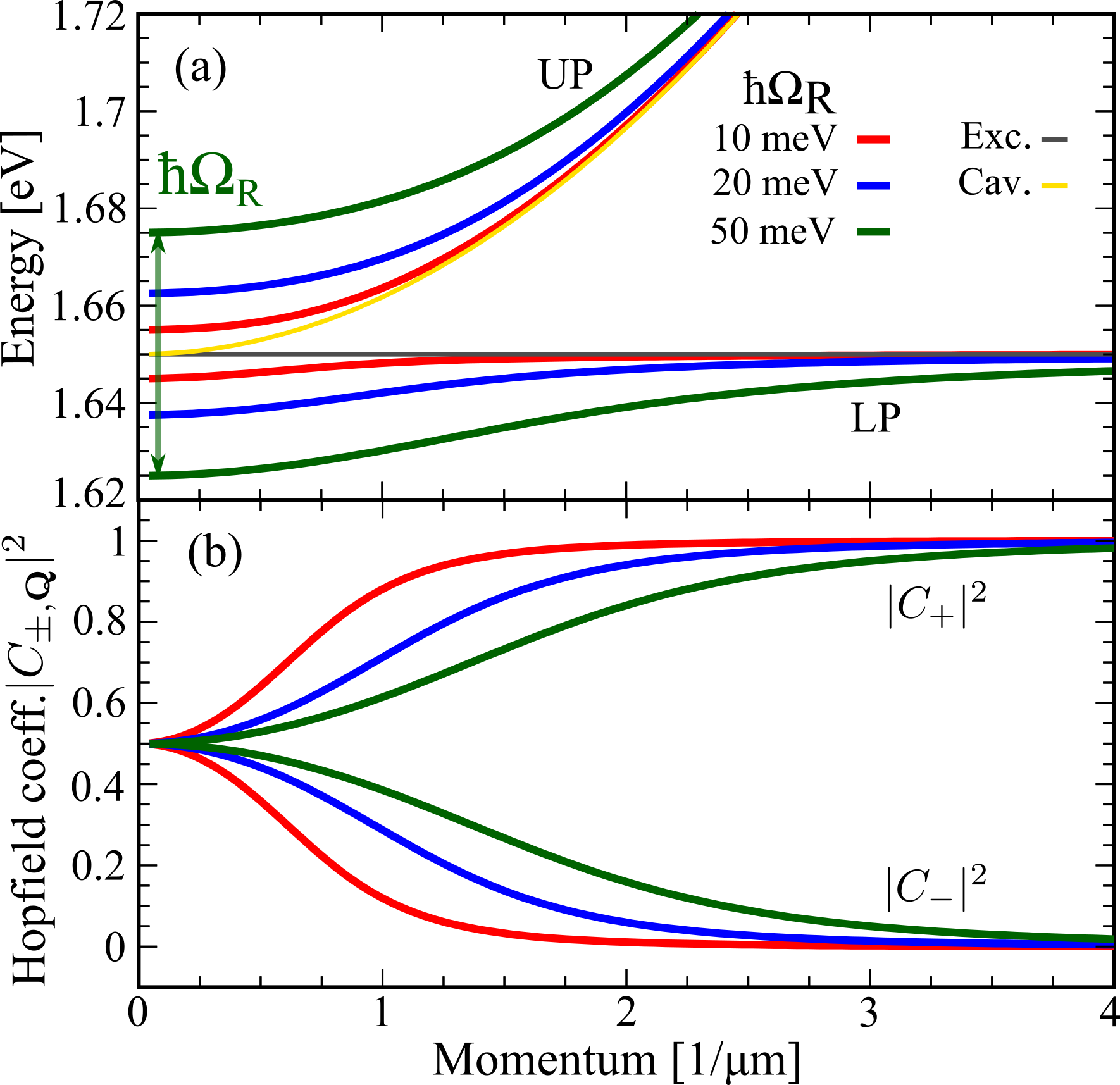}
	\caption{(a) Polariton dispersion and (b) Hopfield coefficients for three typical values of  Rabi splitting. The bare exciton and photon energy  are shown for comparison with thin grey and yellow lines, respectively.}
	\label{fig_1}
\end{figure}

Now, we investigate the strong-coupling regime, where the exciton-photon coupling strength $g_{\mathbf{Q}}$ is larger than (the difference of) cavity and non-radiative exciton decay rates \cite{deng2010exciton}. This allows to form polaritons as new eigenmodes that can be obtained by applying a Hopfield transformation of the Hamilton operator yielding \cite{hopfield1958theory, deng2010exciton}
\begin{equation}
\begin{split}
\hspace{-0.1cm}\hat{H}&\hspace{-0.05cm}=\hspace{-0.1cm}\sum_{\mathbf{Q},i}E^{i}_{\mathbf{Q}}\hat{Y}^{i \dagger}_{\mathbf{Q}}\hat{Y}^{i}_{\mathbf{Q}}+\hspace{-0.3cm}\sum_{\mathbf{Q},\mathbf{q},i,i^\prime}\hspace{-0.25cm}\tilde{\mathcal{D}}^{i^\prime,i}_{\mathbf{q}}\left(\hat{b}^\dagger_{-\mathbf{q}}\hspace{-0.1cm}+\hat{b}_{\mathbf{q}}\right)\hat{Y}^{i^\prime \dagger}_{\mathbf{Q}+\textbf{q}}\hat{Y}^{i}_{\mathbf{Q}}. 
\end{split} \label{hamiltonpol}
\end{equation}
 Here, $\hat Y^\dagger (\hat Y)$ denotes the  polariton creation (annihilation) operator and $E^i_{\mathbf{Q}}$ the energy of the lower and upper polariton branch ($i=$LP,UP), cf. Fig. \ref{fig_1}(a). In many cases, the decay rate of the cavity mode is much larger than the dissipation rate of excitons allowing us to find an analytic expression for the polariton energies $E^{i}_\mathbf{Q}$ reading \cite{deng2010exciton}
\begin{align}
E^{\text{LP/UP}}_\mathbf{Q}&=\frac{1}{2}E^\text{X}_\mathbf{Q}+\frac{1}{2}E^\text{C}_\mathbf{Q}\mp\frac{1}{2}\sqrt{4g_{\mathbf{Q}}^2+\Delta E_\mathbf{Q}^2} \label{Eneg_pol}
\end{align}
with $\Delta E_\mathbf{Q}= E^\text{X}_\mathbf{Q}-E^\text{C}_\mathbf{Q}$.
For a vanishing exciton-phonon coupling $g_{\mathbf{Q}}$, the lower/upper polariton branch approach the cavity and exciton energy (thin yellow and grey lines in Fig. \ref{fig_1}(a)), respectively. Throughout this work, we focus on the resonant case,  i.e. $E^\text{X}_0=E^\text{C}_0$. In the strong coupling regime with a large $g_{\mathbf{Q}}$, an avoided crossing occurs and two polariton branches are formed. Their separation corresponds to the  Rabi splitting $\hbar\Omega_\text{R}=2g_0$. The two polariton branches can be visualized in optical spectra for large-enough coupling $g$ \cite{khitrova2006vacuum,baranov2018novel,hu2020}.

 The polariton states consist of a coherent mixture of excitons and photons with the in-plane momentum $\mathbf{Q}$, thus the polariton operators can be expressed as 
\begin{align}
\hat{Y}^i_\mathbf{Q}=h^{i}_{\text{X},\mathbf{Q}}\hat X_\mathbf{Q}+h^{i}_{\text{C},\mathbf{Q}} \hat{c}_\mathbf{Q} \label{pol_op}
\end{align}
with the Hopfield coefficients $h^{\text{LP}}_{X,\mathbf{Q}}=h^{\text{UP}}_{\text{C},\mathbf{Q}}=C_{+,\mathbf{Q}}$ and $h^{LP}_{c,\mathbf{Q}}=-h^{UP}_{\text{X},\mathbf{Q}}=C_{-,\mathbf{Q}}$ \cite{hopfield1958theory,deng2010exciton}, where 
\begin{align}
|C_{\pm,\mathbf{Q}}|^2=\frac{1}{2}\left(1\pm\Delta E_\mathbf{Q}\left[\Delta E_\mathbf{Q}^2 +4 g_{\mathbf{Q}}^2\right]^{-\frac{1}{2}}\right). \label{hop_coef}
\end{align}

The second  term in the Hamilton operator from Eq. (\ref{hamiltonpol}) provides the polariton-phonon interaction. Since only the excitonic part of the polariton couples with phonons, the appearing matrix element $\tilde{D}$ is related to the exciton-phonon coupling via $\tilde{\mathcal{D}}_{\mathbf{Q}^{\prime}, \mathbf{Q}, \mathbf{q}}^{ i^\prime,i}=h^{i^\prime *}_{X,\mathbf{Q}^\prime}\mathcal{D}_{\mathbf{q}}h^{i}_{X,\mathbf{Q}}$  \cite{Lengers21}. Using the second-order Born-Markov approximation \cite{thranhardt2000,Brem18}, we determine the polariton-phonon scattering rate 
\begin{align}
\Gamma^{i}_{\mathbf{Q}}=&2\pi\sum_{i^\prime}\sum_{\mathbf{Q}^\prime} \,|\tilde{\mathcal{D}}_{\mathbf{Q}^{\prime}, \mathbf{Q}}^{ i^\prime,i }|^2\left(\frac{1}{2}\pm\frac{1}{2}+n_{\mathbf{Q}^\prime-\mathbf{Q}}\right)\nonumber\\
&\times L_{\gamma_{\mathbf{Q}^\prime}}\left(E^{i^\prime}_{\mathbf{Q}^\prime}-E^{i}_\mathbf{Q}\pm\hbar\omega_{\mathbf{Q}^\prime-\mathbf{Q}}\right) \label{pol_sr}
\end{align} 
with the phonon momentum ${\mathbf{q}}=\mathbf{Q}^\prime-\mathbf{Q}$, the Bose-Einstein distribution $n_\mathbf{\mathbf{q}}$, and the phonon energy $\hbar\omega_{\mathbf{q}}$. In our work, we partially include some effects beyond the so-called completed-collision limit \cite{Rossi11} introducing a Lorentzian function instead of a Delta function. This is similar to the damping introduced in the second-order Born approximation including higher-order effects leading e.g. to a collisional broadening \cite{Schilp94,Lengers20abs}.
A microscopic calculation of the broadening $\gamma$ is beyond the scope of this work. Through the manuscript, we use a  value of $0.1$ meV, which provides a low estimation of the presented scattering rates. As further discussed below, larger values of $\gamma$ do not change the qualitative behaviour, but lead to a quantitative increase of the scattering of polaritons with acoustic modes due to higher-order non-resonant contributions.

Having determined the expression for polariton energies and polariton-phonon scattering rates, we have all necessary ingredients to investigate the polariton diffusion. 
Polaritons in general show a peculiar transient spatio-temporal dynamics resulting in a large ballistic propagation even at room temperature \cite{Wurdack21}, as well as a non-linear transport behaviour  \cite{ballarini2013all, kolmakov2016toward}. While a full spatio-temporal polariton dynamics is beyond the scope of this work, we focus the investigation on polariton diffusion coefficients in steady-state limit. In TMD monolayers, excitons show - after an initial unconventional diffusion \cite{Rosati20, Rosati21c}-  a regular steady-state diffusion behaviour, i.e. exhibiting a linear increase of the square width of the spatial distribution as a function of time \cite{Kumar2014t,Yuan17,Kato16,Cadiz17,Kulig18}. The rate of this increase is given by the diffusion coefficient $D$ \cite{Hess96, Rosati21c}.
Such a regime appears when a local thermalized distribution is reached and when the scattering processes are fast enough leading to a quick thermalization compared to the transport timescale \cite{Hess96}. The diffusion coefficient $D$ can be obtained from the general definition of current $\mathbf{J}(\mathbf{r})=-D \nabla_\mathbf{r} N(\mathbf{r})$ with $N(r)\propto\sum_{\mathbf{Q}} f(\mathbf{Q},\mathbf{r},t)$ being the spatial distribution, where $f(\mathbf{Q},\mathbf{r},t)$ is the Wigner function.  
Assuming fast exciton-phonon scattering  and quasi-thermalized local distributions \cite{Hess96}, an analytic expression for the diffusion coefficient can be obtained \cite{Hess96}
\begin{align}
D=\frac{\hbar}{2}\sum_{\mathbf{Q},i}\frac{|v_\mathbf{Q}^i|^2}{\Gamma^{i}_{\mathbf{Q}}}f^i_\mathbf{Q} \quad , \label{pol_diff}
\end{align}
with the polariton group velocity $v_\mathbf{Q}^i$ and occupation probability $f^i_\mathbf{Q}$. The latter is assumed to be a thermalized Boltzmann distribution $f^i_\mathbf{Q}\propto e^{-E^i_\mathbf{Q}/k_B T}$ in the low-density limit (with $k_B$ as the Boltzmann constant and $T$ as the temperature). Note that this approximation can in general lead to an overestimation of the actual occupation of  quasi-photonic polariton states in the upper polariton branch. However, in the considered regime of resonant exciton and cavity energies, not too high temperatures  and relatively large  Rabi splittings, these effects are found to be small.
According to Eq. (\ref{pol_diff}), the crucial quantities determining the polariton diffusion are the polariton group velocity $v_\mathbf{Q}^i$, the polariton-phonon scattering rate $\Gamma^{i}_{\mathbf{Q}}$ and the occupation of polariton states $f^i_\mathbf{Q}$. 

The many-particle mechanisms behind the diffusion can differ considerably when moving from TMD monolayers to TMD bulk materials. In the monolayer case, the reduced screening leads to large excitonic binding energies. As a consequence, the diffusion is typically dominated by excitons. Nevertheless, contributions from the faster diffusing electron-hole plasma can still appear for substrates with a large dielectric constant, as observed  for hBN-encapsulated TMDs at higher temperatures \cite{zipfel2020}. Bulk materials are expected to have smaller excitonic effects and thus higher diffusion coefficients in the range of 10 cm$^2$/s, as observed e.g. for MoS$_2$ \cite{kumar2014} and MoTe$_2$ \cite{pan2018}. 
In the next section, we discuss each of these microscopic quantities before we analyze the polariton diffusion coefficients and their temperature dependence.

%%%%%%%%%%%%%%%%%%%%%%%%%%%%%%%%%%%%%%%%%%%%%%%%%%%%%%%%%%%%%%%%%%%%%%%%%%%%%%%%%%%%%%%%%%%%%%%%%%%%%%%%%%%%%%%%%%%%%%%%%%%%%%%%%%%%%%%%
\section{Results}\label{sec3}
\subsection{Polariton group velocity and occupation}
Now, we investigate the change of the excitonic band structure in the presence of a strong coupling regime. Figure \ref{fig_1}(a) illustrates the polariton dispersion  for three  values of the   Rabi splitting representing different exciton-photon coupling strengths. The latter depends on the oscillator strength of the material and the characteristics of the  optical cavity. 
Exciton-photon coupling induces a Rabi splitting $\hbar \Omega_\text{R}= E^\text{UP}_0-E^\text{LP}_0 =2g_0$ and the formation of an upper and a lower polariton branch.  We investigate the polariton dispersion for typical  Rabi splitting values of $\hbar \Omega_\text{R}=$10, 25, 50 meV \cite{schneider2018}, which are  larger than the non-radiative exciton linewidth (typically a few meV at low temperatures for hBN-encapsulation TMDs \cite{Selig16,Cadiz17b,Christiansen17}) and the cavity linewidth (ranging from the meV \cite{dufferwiel2015} down to the $\mu$eV range \cite{steger2013,delteil2019}), thus allowing for strong-coupling regime \cite{deng2010exciton}.

At larger momenta, the LP and UP branches merge with the exciton and photon dispersion, respectively. The larger $\hbar \Omega_\text{R}$, the higher are the momentum values at which this occurs, cf. Fig. \ref{fig_1}(a). 
Polaritons are coherent superpositions of excitonic and photonic states with the Hopfield coefficients defining the weights of the single constituents, cf.  Fig. \ref{fig_1}(b). The coefficient $\vert C_{+,\mathbf{Q}}\vert^2$ gives the exciton content of the lower polariton and the photon content of the upper polariton, i.e. for $\vert C_{+,\mathbf{Q}}\vert^2=$1 the LP state $\vert \text{LP}, \mathbf{Q} \rangle=\hat{Y}^{\text{LP}\dagger}_\mathbf{Q} \vert 0 \rangle$ coincides with the exciton state $\vert \text{X},\mathbf{Q} \rangle$ and the UP state $\vert \text{UP},\mathbf{Q} \rangle$ corresponds to the photon state $\vert \text{C},\mathbf{Q} \rangle$.

\begin{figure}[t!]
\centering
	\includegraphics[width=\columnwidth]{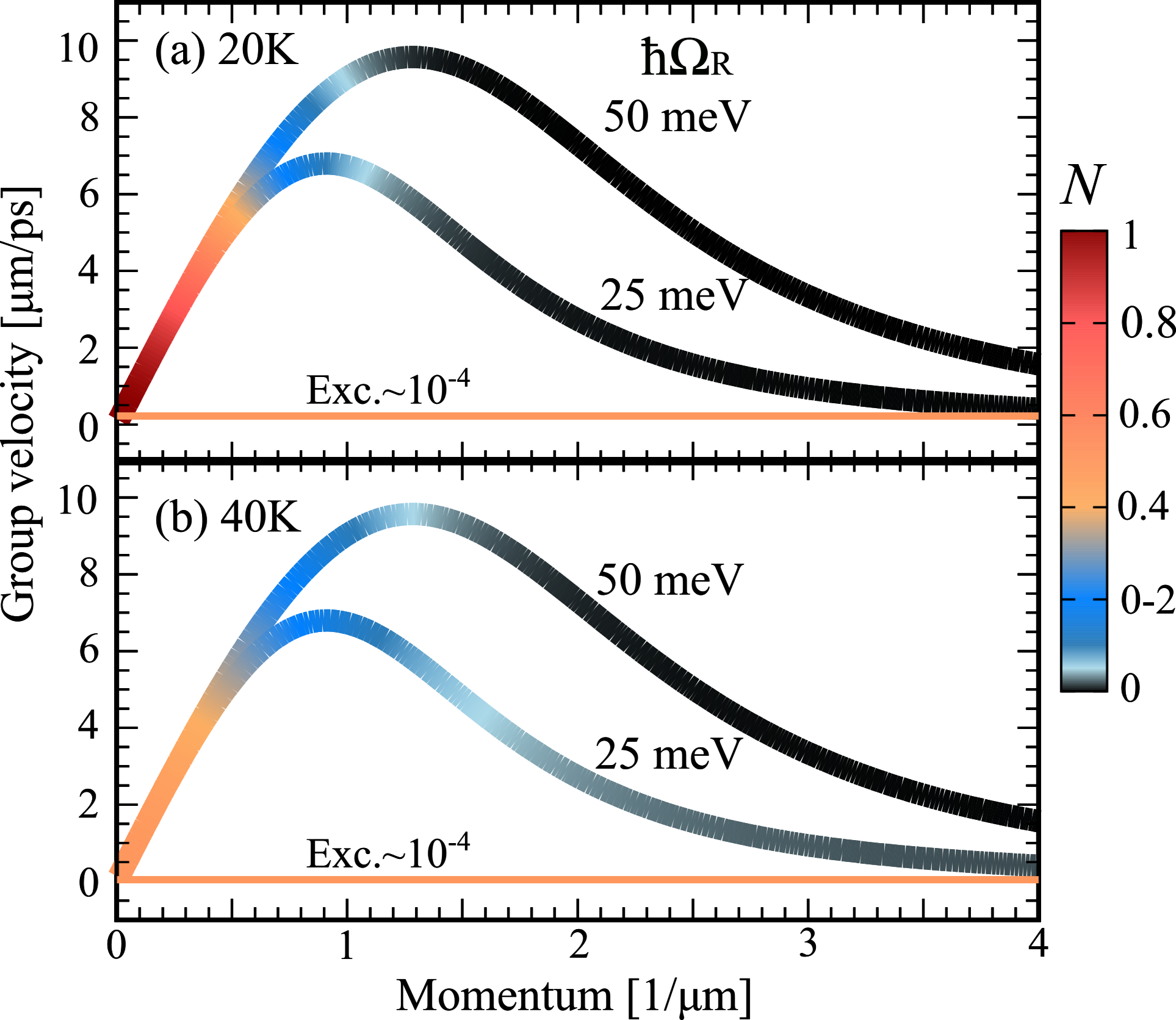}
	\caption{Lower polariton group velocity  at (a) $T$=20 K and (b) $T$=40 K  for two different  Rabi splittings of $\hbar \Omega_R=25$ and 50 meV. The corresponding band occupation is overlaid on the polariton dispersion, see the colour map. For comparison, the excitonic group velocity is shown by the thin orange line. }
	\label{fig_2}
\end{figure}

In Fig. \ref{fig_2}, we consider the polariton group-velocity 
$v^i_Q=\hbar^{-1}dE^i_Q/dQ$, in particular focusing on the lower polariton branch, since the upper one has a negligible occupation and thus its impact on diffusion will be limited.  
We consider the two cases of  Rabi splitting $\hbar \Omega_\text{R}$=25, 50 meV in comparison with the excitonic group velocity $v^{X}_Q=\hbar Q/M_X$, where $M_X$ is the exciton mass.
We find that the polariton group velocity is approximately 4 to 5 orders of magnitude larger than the excitonic one for small momenta within the light cone, cf. Fig. \ref{fig_2}. Due to the rapidly changing polariton dispersion, we find group velocities in the range of 10 $\mu$m/ps, thus principally opening the possibility of ballistic polariton propagation for 10s $\mu$m. This has recently indeed been observed in a space- and angle-resolved photoluminescence experiments on a WS$_2$ monolayer in a distributed Bragg reflector cavity \cite{Wurdack21}.
In addition to the remarkable magnitude difference, the group velocity for polaritons has also a  qualitatively different momentum dependence. It shows a maximum in correspondence to the inflection point in the lower polariton branch and decreases toward the excitonic velocity for momenta of several $\mu\text{m}^{-1}$. 

In a nutshell, two subset of states with a considerably different group velocity coexist in a cavity: The fast ones located within the light cone and the slow ones coinciding with conventional excitons. However, a large  group velocity alone is not enough to boost the increase of diffusion coefficients, but these states also need to be occupied. This enters through the momentum-dependent Boltzmann distribution in Eq. (\ref{pol_diff}). To illustrate this, we overlay the occupation of the lower polariton state on the line displaying its group velocity (reddish colours denote large occupation), cf. Fig. \ref{fig_2}. While the excitonic occupation is momentum-independent in the considered range of momenta (cf. the thin orange line in Fig. \ref{fig_2}), strong variations are observed for polaritons. At 20 K, the occupation of the states at larger momenta  is decreased by two orders of magnitude with respect to the exciton case for both considered Rabi splittings ($f^\text{LP}/f^\text{X} \approx 10^{-5}$ at $Q\approx$4$\mu\text{m}^{-1}$), cf.  orange vs black colour in Fig. \ref{fig_2}. The curvature of the polariton branch induces a significant decrease of the occupation of the slow quasi-excitonic states at large momenta, as the energetically lower states at $Q\approx$0 are more efficiently populated. 
Increasing the temperature, the population of the former starts to increase, in particular for the smaller Rabi splitting of 25 meV, cf. Fig. \ref{fig_2}(b). Regarding the behaviour at smaller momenta, we see that at 20 K the occupation of states with the maximum group velocity is negligible for $\hbar \Omega_\text{R}$=50 meV (black colour at approximately $Q\approx1.3 \mu$m$^{-1}$). However, when increasing the temperature to  40 K, we find a considerable occupation even at these states indicating the possibility of a strongly accelerated polariton diffusion.

\subsection{Polariton-phonon scattering}
\begin{figure}[t!]
\centering
	\includegraphics[width=\columnwidth]{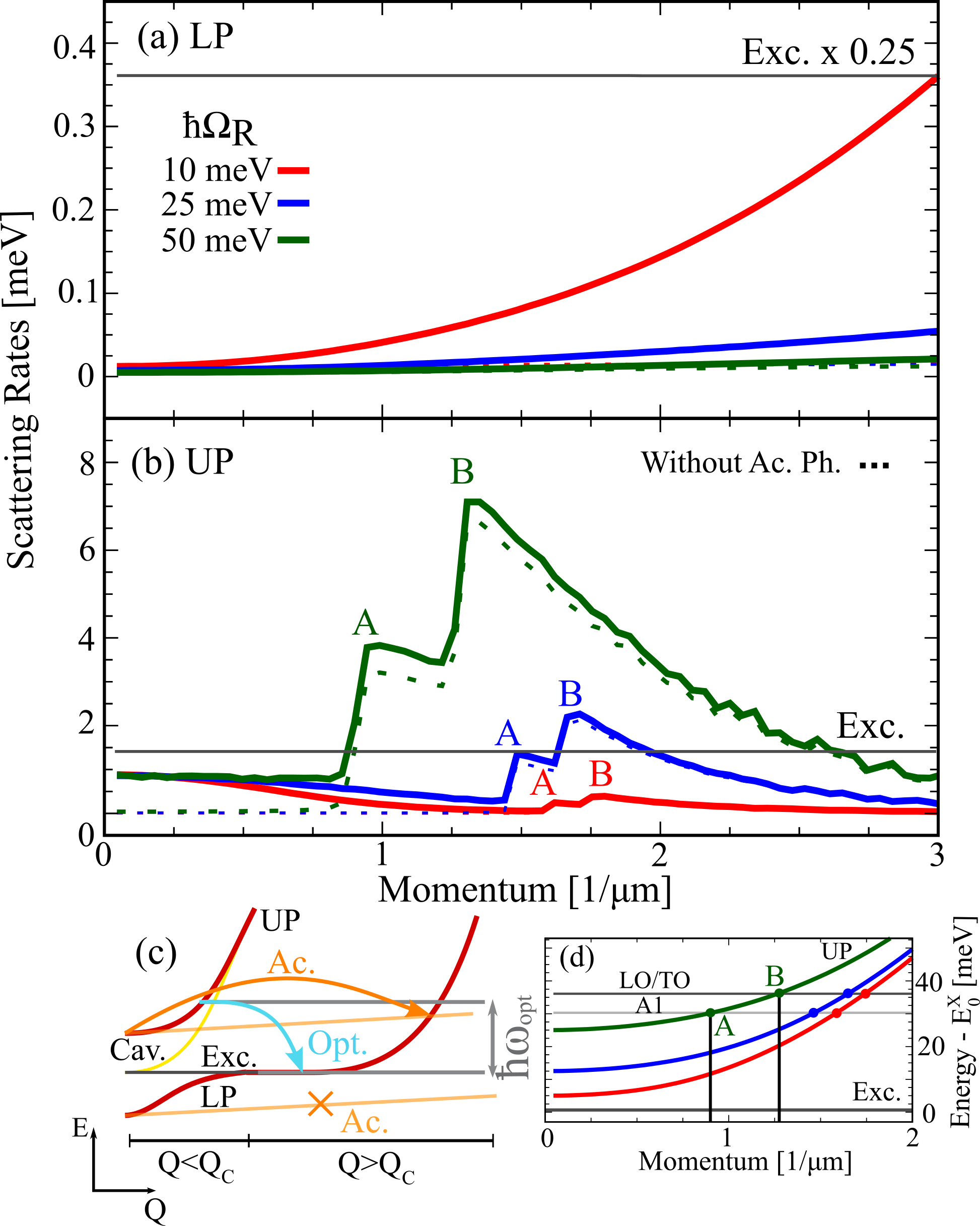}
    \caption{Polariton-phonon scattering rates  at $T=40$ K for (a) the lower  and (b) the upper polariton branch. Dashed lines show the case without the contribution of acoustic phonons. Note that in (a) all three dashed lines lie close to each other and are nearly momentum-independent illustrating the crucial role of acoustic phonons for the increased scattering at large momenta. (c) Schematic representation of possible scattering channels induced by emission and absorption of acoustic and optical phonons. (d) Crossing points of optical phonon dispersion with the upper polariton branch for different $\hbar \Omega_R$ explaining the resonances appearing in Fig. \ref{fig_3}(b).}
    \label{fig_3}
\end{figure}

Besides the group velocity and occupation of polariton states, polariton-phonon scattering plays an important role for the diffusion coefficient (cf. Eq. \ref{pol_diff}). 
Here, we neglect scattering with defects/disorder \cite{Wurdack21} or intervalley scattering with KK$^\prime$ excitons.
Figure \ref{fig_3} illustrates the polariton-phonon scattering rate at 40 K (cf.  Eq. (\ref{pol_sr}) for LP and UP branches around the light cone). Note that only the scattering into LP states out the light cone is efficient  due to the limited number of receiving partner states available within the light cone as well as due to the negligibly small Hopfield coefficients $h^{\text{UP}}_{\text{X}}$ for large-momenta UP states. This implies that the receiving LP state is 
quasi-excitonic, cf. Fig. \ref{fig_1}, i.e. the associated coefficient fulfills $h^{\text{LP}}_{\text{X}}\approx 1$ and the related scattering coefficient are proportional to $\left|\tilde{\mathcal{D}}_{\mathbf{Q}^{\prime}, \mathbf{Q}}^{ \text{LP},i}\right|^2\approx \left|\mathcal{D}_{\mathbf{Q}^\prime-\mathbf{Q}}\right|^2\left|h^{i}_{X,\mathbf{Q}}\right|^2$. 
As a result, 
the scattering with phonons 
is driven by the excitonic component of the emitting polariton. One would naively expect larger scattering rates for LP states, as here the excitonic constituent is dominant reflected by $\left\vert h^\text{LP}_X\right\vert^2\equiv \left\vert C_+\right\vert^2 \approx 1$. Surprisingly, our microscopic calculations of scattering rates  show a much more efficient scattering for the UP branch, cf. Fig. \ref{fig_3}. This can be traced back to the number of available scattering states fulfilling the momentum and energy conservation. 

First, we discuss the LP scattering rates shown in Fig. \ref{fig_3}(a). 
For states around $Q$=0, we find two orders of magnitude smaller polariton-phonon scattering compared to the exciton case (thin black line), while the Hopfield coefficient $\left\vert C_+\right\vert^{2}=0.5$ would only imply a decrease by a factor of two.
The reason for the dramatic decrease is related to the change in the dispersion relation in the strong coupling regime. 
The energy of acoustic phonons $\hbar v_s Q$ is almost flat compared to polaritons (cf. orange and red line in Fig. \ref{fig_3}(c), respectively). As a consequence, when low-momentum polaritons absorb acoustic phonons, they are not able to find a resonant scattering partner. This is only possible if they are very close  to the exciton energy $E^X_{0}$.% (concretely, the separation has to be smaller than $\Delta_E\lesssim 0.05$ meV reflecting the resonance condition  $E^X_{0}-\Delta_E+\hbar v_sQ=E^X_{0}+\hbar^2Q^2/2M$). 
The increase of the scattering rate at larger momenta can be traced back to  non-resonant scattering with acoustic phonons (cf. the dashed line in Fig. \ref{fig_3}(a) excluding acoustic phonons). This is due to the width of the Lorentzian in Eq. \ref{pol_sr}, whose origin can be related to higher-order scattering contributions inducing a softening of the energy selection rules. Note that the scattering rates show a quantitative dependence on  this phenomenologically introduced width parameter, however the qualitative behaviour remains unaffected. 

Next, we discuss the UP scattering rates illustrated in Fig. \ref{fig_3}(b). At very small momenta, the scattering with acoustic modes is much more efficient and reaches a value that is approximately only two times smaller than for excitons, as expected by the Hopfield coefficient (Fig. \ref{fig_1}(b)). Since the UP branch is higher in energy compared to excitons,  it is possible to find resonant scattering partners, cf. the crossing between the phonon dispersion (top orange line) and LP energy in Fig. \ref{fig_3} (c)). As a result, the scattering  via acoustic phonons is efficient. At larger momenta, we observe the appearance of pronounced A and B resonances reflecting the emission of  optical phonons.  To better understand their origin, in Fig. \ref{fig_3}(d) we plot the UP dispersion and the optical phonon energies (with respect to the exciton energy $E^X_{0}$) and find crossing points exactly at the position of the A and B peaks in the scattering rate.
Note that the different weight of these peaks for different Rabi splitting is due to the Hopfield coefficients. For $\hbar \Omega\text{R}$=50 meV, the A peak appears at $Q\approx 1 \mu$m$^{-1}$, where $\vert C_+\vert^2\approx 0.4$, while at 10 meV it appears at $Q\approx 1.5 \mu$ m$^{-1}$ with $\vert C_+\vert^2\approx 0.04$ resulting in a much smaller scattering efficiency.
The UP scattering rates are dominated by resonant scattering, hence the width of the Lorentzian plays a minor role.

In a nutshell,  the polariton-phonon scattering is strongly affected by the polariton dispersion resulting in  suppressed  scattering with acoustic phonons for LP and an enhanced emission of optical phonons for UP states. Note that the phonon-induced scattering also contributes to the polariton linewidth in optical spectra \cite{Lengers21,dufferwiel2015}. Here, angle-resolved spectroscopy reaching states with non-zero momenta \cite{liu2015} could principally allow to  measure the predicted strongly momentum-dependent scattering rates via a change in the polariton linewidth, in particular  for the UP branch.

\begin{figure}[t!]
\centering
	\includegraphics[width=0.44\textwidth]{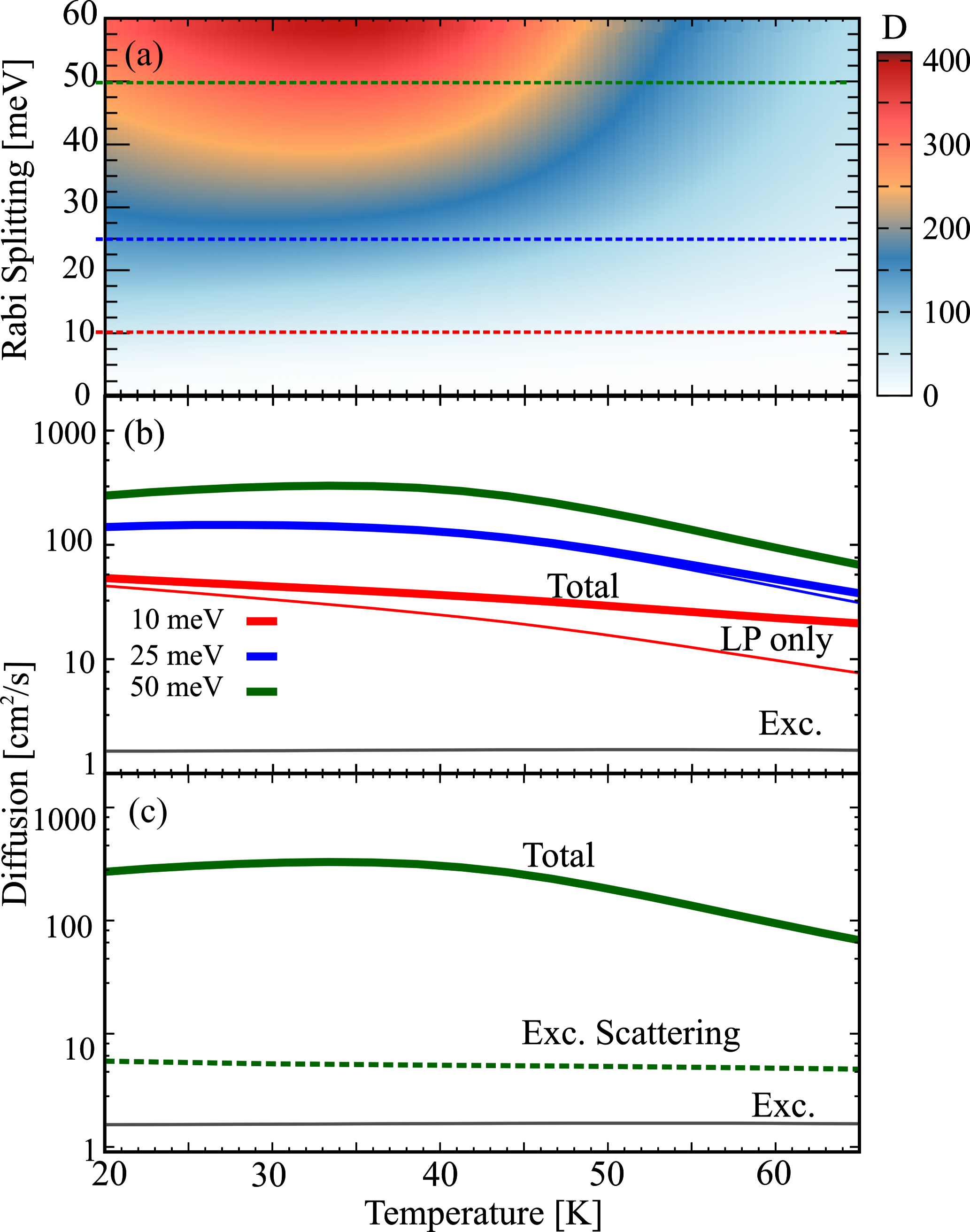}
    \caption{(a) Polariton diffusion as a function of temperature and Rabi splitting with (b) cuts evaluated for fixed splittings of 10, 25 and 50 meV. Here, thin lines show the  diffusion 
     stemming only from lower polaritons. (c) Polariton diffusion assuming exciton-phonon scattering rates  illustrating the impact of the changed scattering for polaritons.}
    \label{fig_4}
\end{figure}

\subsection{Polariton diffusion}
We have now discussed all key ingredients to evaluate and understand the polariton diffusion. 
Figure \ref{fig_4}(a) shows the diffusion coefficient as a function of temperature and  Rabi splitting. Based on our microscopic approach, we predict polariton diffusion coefficients that are two to three orders of magnitude larger than the ones from  the bare exciton. This can be explained by: (i) the polariton dispersion exhibiting huge group velocities, (ii) effective occupation of fast polaritonic states, and (iii) reduced scattering with phonons of the occupied lower polariton states.
These features concern, however, only small-momentum polaritons, while the diffusion coefficients depend also on large-momentum states. The latter are unaffected by point (i) and (iii), as the polariton dispersion and scattering rates correspond to the excitonic values at large momenta (Figs. \ref{fig_2}, \ref{fig_3}). Only the relative population remains affected by the presence of lower lying polariton states at small momenta.  As a result, the polariton diffusion coefficient is the result of a non-trivial interplay between the few very fast states within the light-cone and  excitonic-like states outside of the cone.
Interestingly, for $\hbar\Omega_\text{R}=50$ meV we observe a maximum in the polariton diffusion at around 40K. This can be traced back to  the occupation discussed in Fig. \ref{fig_2}: At 40 K, fast polariton states with a maximum group velocity at approximately  $Q\lesssim 1.3 \mu$m$^{-1}$ are efficiently populated resulting in a maximum diffusion. Further increasing the temperature occupies states at higher momenta (and a smaller group velocity)  inducing a decrease of $D$.

In Fig. \ref{fig_4}(b), we show the temperature dependence of the polariton diffusion coefficient at three fixed value for the  Rabi splitting in comparison with the excitonic value (thin grey line). For increasing temperature, the polariton diffusion decreases towards the bare exciton diffusion (with $D$ in the range of a few cm$^2$/s) and this occurs faster for smaller Rabi splittings $\hbar\Omega_\text{R}$. At higher temperatures, the amount of occupied slower quasi-excitonic states outside the light cone becomes larger. 
Comparing the total diffusion coefficient with the contribution stemming only from  the lower polariton states (thin lines), we find that  the LP contribution is dominant for $\hbar \Omega_\text{R}=$ 25 and 50 meV, as the UP states are only marginally occupied for the considered low temperatures. 
For the lower Rabi splitting of $\hbar\Omega_\text{R}$=10 meV, we find that the total and the LP  diffusion start to deviate at higher temperatures indicating the increasing weight of the UP diffusion.  \\
To illustrate  the impact of the polariton-phonon scattering on the diffusion, we calculate in a gedanken experiment the polariton diffusion assuming exciton scattering rates, cf. the dashed line in  Fig.\ref{fig_4}(c). We observe a significant decrease in the polariton diffusion by more than one order of magnitude indicating the important role of polariton-phonon scattering.
Note however that the polariton diffusion coefficient still remains considerably higher than the exciton one (thin grey line) reflecting the strong impact of the polariton group velocity. Interestingly, the peak at 40 K shown in the full polaritonic case disappears as the excitonic scattering rates decrease the distinction between fast and slow polariton states in view of their weak momentum dependence (Fig. \ref{fig_3}(a)).\vfill\null

%%%%%%%%%%%%%%%%%%%%%%%%%%%%%%%%%%%%%%%%%%%%%%%%%%%%%%%%%%%%%%%%%%%%%%%%
%%%%%%%%%%%%%%%%%%%%%%%%%%%%%%%%%%%%%%%%%%%%%%%%%%%%%%%%%%%%%%%%%%%%%%%%

\section{Conclusion}\label{sec4}
We have microscopically calculated the polariton diffusion coefficient for an hBN-encapsulated MoSe$_2$ monolayer embedded in an optical cavity.
We predict a drastically enhanced polariton diffusion that is two to three orders of magnitude higher compared to excitons. We show that this accelerated polariton diffusion can be traced back to the increased group velocity of polaritons and the decreased scattering rates with phonons reflecting the dual light-matter character of polaritons. 
\\
\section*{Acknowledgments}
This project has received funding support from the DFG via SFB 1083 (project B9), the European Union's Horizon 2020 Research and Innovation programme under grant agreement no. 881603 (Graphene Flagship) and from the Knut and Alice Wallenberg Foundation via the Grant KAW 2019.0140. The computations were enabled by resources provided by the Swedish National Infrastructure for Computing (SNIC).
We thank Jamie Fitzgerald (Chalmers) for valuable discussions that helped us in developing this work. \\

\end{document}